\documentclass[11pt]{article}
\usepackage[margin=1in]{geometry}
\usepackage{graphicx}
\usepackage{natbib}
\usepackage{footnote,enumerate,amsmath,amssymb,amsfonts,amsthm,subcaption,hyperref}

\newcommand{\N}{{\mathbb N}}

\begin{document}
\title{Reducing Binary Quadratic Forms for More Scalable Quantum Annealing}
\author{Georg Hahn\\Imperial College, London, UK
\and Hristo Djidjev (PI)\\Los Alamos National Laboratory}
\date{}
\maketitle

\begin{abstract}
Recent advances in the development of commercial quantum annealers such as the D-Wave 2X allow solving NP-hard optimization problems that can be expressed as quadratic unconstrained binary programs. However, the relatively small number of available qubits (around 1000 for the D-Wave 2X quantum annealer) poses a severe limitation to the range of problems that can be solved. This paper explores the suitability of preprocessing methods  for reducing the sizes of the input programs and thereby the number of qubits required for their solution on quantum computers. Such methods allow us to determine the value of certain variables that hold in either any optimal solution (called strong persistencies) or in at least one optimal solution (weak persistencies). We investigate preprocessing methods for two important NP-hard graph problems, the computation of a maximum clique and a maximum cut in a graph. We show that the identification of strong and weak persistencies for those two optimization problems is very instance-specific, but can lead to substantial reductions in the number of variables.
\end{abstract}

\section{Introduction}
\label{sec:intro}

\subsection{Quantum annealing and D-Wave}
The recent availability of the first commercial quantum computers of D-Wave Systems Inc.~\citep{dwave2016} provides a novel tool to help finding,  by a process known as \textit{quantum annealing}, global solutions to NP-hard problems that seem to be  difficult to obtain classically~\citep{Djidjev2017}.

As of 2017, the most recent version of the company's quantum annealer, D-Wave 2000Q, is capable of handling around $2000$ \textit{qubits} of quantum information. Each qubit consists of a superconducting loop on the D-Wave chip which simultaneously encodes a quantum superposition of $-1$ and $+1$ via two superimposed currents in both clockwise and counter-clockwise directions~\citep{Johnson2011,Bunyk2014}. Upon completion of the annealing process, the system turns classical and each qubit takes a classical (binary) value that can be read and assigned to a variable. D-Wave is designed to minimize an  objective function consisting of a sum of linear and quadratic binary contributions, given as
\begin{equation}
  H=H(x_1,\dots,x_N)=\sum_{i \in V} a_i x_i + \sum_{(i,j) \in E} a_{ij} x_i x_j, \label{eq:hamilt}
\end{equation}
where $a_i$, $a_{ij} \in \mathbb{R}$  and $V = \{1,\ldots,N\}$, $E=V \times V$ for $N \in \N$ qubits~\citep{King2015}. The functional type \eqref{eq:hamilt} is called the \textit{Hamiltonian} encoding the  energy of a quantum system. In \eqref{eq:hamilt} qubits are encoded by variables $x_i$, $E$ denotes the dependency structure between the qubits (the set of interactions), and $a_i$ and $a_{ij}$ encode physical characteristics of the qubits and the links (called \textit{couplers}) between them, respectively. During annealing, the quantum system tries to reach a configuration of minimal energy of $H$. A quadratic form of type \eqref{eq:hamilt} with $x_i \in \{-1,1\}$ is called an \textit{Ising model}. By replacing $x_i$ in \eqref{eq:hamilt} with $(x_i+1)/2$, one gets a quadratic form of the same type as \eqref{eq:hamilt}, but with $x_i \in \{0,1\}$. The resulting problem is called a \textit{quadratic unconstrained binary optimization (QUBO)} problem. Any QUBO or Ising problem can be solved on D-Wave as long as there are enough qubits and couplers between those qubits available on the D-Wave architecture to represent the system \eqref{eq:hamilt}.

\subsection{Aim of this work}
Although \cite{Djidjev2017} show that certain instances designed to fit the D-Wave qubit interconnection network can be solved magnitudes faster with D-Wave than with its classical analogues, the limitation of only around $1000$ qubits and about 3000 couplers available on D-Wave 2X (DW) at Los Alamos National Laboratory  poses a severe restriction. For most NP-hard optimization problems, due to the density of the Hamiltonian $H$, the largest problems able to fit DW are of size about only 45. 

To mitigate the current size limitations of the D-Wave architecture, this work tries to explore the usability of preprocessing methods allowing to determine the value of certain variables in a QUBO or Ising problem. Fixing all such variables then results in a new Hamiltonian of reduced size, consisting only of the unresolved variables. We investigate the usage of partitioning and roof duality methods for two important NP-hard graph problems, the Maximum Clique and the Maximum Cut problem, which have multiple applications including network analysis, bioinformatics, and computational chemistry. Given an undirected graph $G = (V, E)$, a \textit{clique} is a subset $S$ of the vertices forming a complete subgraph, meaning that any two vertices of $S$ are connected by an edge in $G$. The clique size is the number of vertices in $S$, and the \textit{Maximum Clique} problem is to find a clique in $G$ with a maximum number of vertices \citep{Balas1986}. A cut $S$ of a graph $G$ is the set of (cut) edges between $S$ and $V \setminus S$ in $G$. The \textit{Maximum Cut} problem is to find a maximum cut, which is a cut of maximum size.

\subsection{Previous work}
In \cite{BorosHammer2001},
optimization methods for quadratic \textit{pseudo-boolean} functions, which are multi-linear polynomial functions of boolean variables, are considered. The authors analyze a variety of techniques based on roof duality, introduced in \cite{Hammer1984}, in order to determine partial assignments of values or relations between variables in such pseudo-boolean functions. A comprehensive overview of efficient preprocessing techniques for QUBO optimization can be found in \cite{Boros2006}: Those techniques consists of methods based on first and second order derivatives, roof duality and probing techniques resulting in lower bounds on the minimum energy of a QUBO, the values in either any (strong persistencies) or at least one optimum (weak persistencies), binary relations between variables in some or every optimum, as well as decomposition techniques of the original problem into several smaller pairwise independent QUBO subproblems.

A more efficient implementation of the probing techniques of \cite{Boros2006}, combined with the aforementioned roof duality, is presented in \cite{Rother2007}, with a special focus on computer vision applications. The code for the methods investigated in \cite{Rother2007} are made available for \textit{C++} in the \textit{Qpbo} package of \cite{Kolmogorov2014}. Bindings of this package for \textit{Python} are available (package \textit{PyQpbo}, which is part of the \textit{PyStruct} package of \cite{JMLR:v15:mueller14a}) and will be used in the experiments of this article.

The article is organized as follows. Section~\ref{sec:background} gives a detailed overview of the optimization methods used in this report. Moreover, details on the optimization problems for the two NP-hard graph problems we consider, Maximum Clique and Maximum Cut, are given. Section~\ref{sec:maxclique} focuses on the Maximum Clique problem. We state its formulation as a QUBO, highlight the use of a decomposition method based on graph partitioning methods into smaller subproblems, and present experimental results for Qpbo. For this we employ several test graphs, both stand-alone and in connection with the decomposition approach. Section~\ref{sec:maxcut} repeats the QUBO formulation and a Qpbo experimental study for the Maximum Cut problem. We conclude with a discussion in Section~\ref{sec:conclusions}.

In the rest of the paper, we denote a graph as $G=(V,E)$, where $V = \{ 1,\ldots,n \}$ is a set of $n$ vertices and $E$ is a set of undirected edges.

\section{Background}
\label{sec:background}
We briefly overview the techniques used in \cite{Boros2006} and \cite{Rother2007} in order to identify strong and weak persistencies. Given a QUBO instance, it is first represented as a (quadratic) \textit{posiform} \begin{align} \phi(x) = \sum_{T \subseteq {\bf L}} a_T \prod_{u \in T} u, \label{eq:posiform} \end{align} where $a_T \geq 0$ for all $T \neq \emptyset$ and $\bf L$ is a set of literals over the $n$ variables $\{1,\ldots,n\}$ and their complements. It is shown in \cite{Boros2006} that any posiform \eqref{eq:posiform} can be equivalently expressed as a quadratic form $$\Phi(x) = a_0 + \sum_{u \in {\bf L}} a_u u + \sum_{u,v \in {\bf L}} a_{uv} u v.$$ Moreover, \cite{Boros2006} shows that there is a one-to-one correspondence between posiforms and \textit{implication networks}, which are capacitated networks $G_\Phi=(V,E)$ with $n=|V|$ nodes and non-negative capacities $\gamma_{uv}$ on the edges connecting any two nodes $u,v \in V$, $u \neq v$, $V=\{1,\ldots,n\}$.

The implication network $G_\Phi$ has an important application in the computation of the \textit{roof dual}, which is a lower bound on the value of a quadratic posiform \citep{Hammer1984}. The roof dual is computed with the help of a maximum flow through the implication network $G_\Phi$.

More importantly for our work, the flow analysis of the implication network also identifies strong and weak persistencies for problem \eqref{eq:posiform}, which are then transformed for persistencies for problem \eqref{eq:hamilt}.

As shown in \cite{Boros2006}, the effectiveness of the roof duality method for finding persistencies can be improved by using \textit{probing}, which is a method that assigns values of $0$ and $1$ to some variables and then compares any persistencies found for these alternative assignments. In some cases, such probing analysis can (a) determine additional persistencies for the original problem or (b) identify relationships between variables which must hold in a (local) optimum, meaning that some pairs of variables will have equal or inverted values. This then allows to substitute one of the variables in each pair with the other (or its inverse).

\section{The maximum clique problem}
\label{sec:maxclique}

\subsection{QUBO formulations}
\label{sec:maxclique_qubo}
We are interested in finding a maximum clique of a graph $G=(V,E)$, where $V=\{1,\ldots,n\}$. We assign a decision variable $x_i$ to each vertex $i$, $i \in \{1,\ldots,n\}$, indicating whether or not vertex $i$ is part of the maximum clique. One way to formulate the Maximum Clique problem as a QUBO is to use the equivalence between Maximum Clique and the maximum independent set problem. The latter asks for a maximum \textit{independent set} of vertices of $G$, that is a set such that no two vertices from it are connected by an edge in $G$. Specifically, an independent set of $H = (V, \overline{E})$ defines a clique in graph $G=(V,E)$, where $\overline{E}$ is the complement of set $E$. This leads to a constrained formulation of Maximum Clique given by
\begin{align}
\underset{x_v \in \{0,1\}}{\text{maximize}}~\sum_{v \in V} x_v \qquad \text{subject to}~\sum_{(u,v) \in \overline{E}} x_u x_v=0.
\label{eq:mis}
\end{align}
The constraint formulation can equivalently be written in QUBO form as
\begin{align}
H = -A\sum_{v \in V} x_v + B\sum_{(u,v) \in \overline{E}} x_u x_v,
\label{eq:maxclique_acm}
\end{align}
where the penalty weights can be chosen as $A=1$, $B=2$ \citep{Lucas2014}. One disadvantage of \eqref{eq:maxclique_acm} lies in the fact that $H$ contains $O(n^2)$ quadratic terms for sparse graphs (of $O(n)$ edges), limiting the size of problems that can fit DW.

An alternative QUBO formulation of \cite{Lucas2014} assumes the clique size $K \geq 1$ is known. Its Hamiltonian
\begin{align}
H_K = A \bigg( K - \sum_{v \in V} x_v \bigg)^2 + B\bigg(\binom{K}{2}  - \sum_{(u,v) \in E} x_u x_v \bigg)
\label{eq:maxclique_lucas}
\end{align}
is designed to attain the value zero only for an assignment defining a clique of size $K$, which is achieved if one chooses weights $A=K+1$ and $B=1$  \citep{Lucas2014}.

\subsection{Problem decomposition}
\label{sec:decomp}

\begin{figure}
  \centering
  \includegraphics[width=0.5\textwidth]{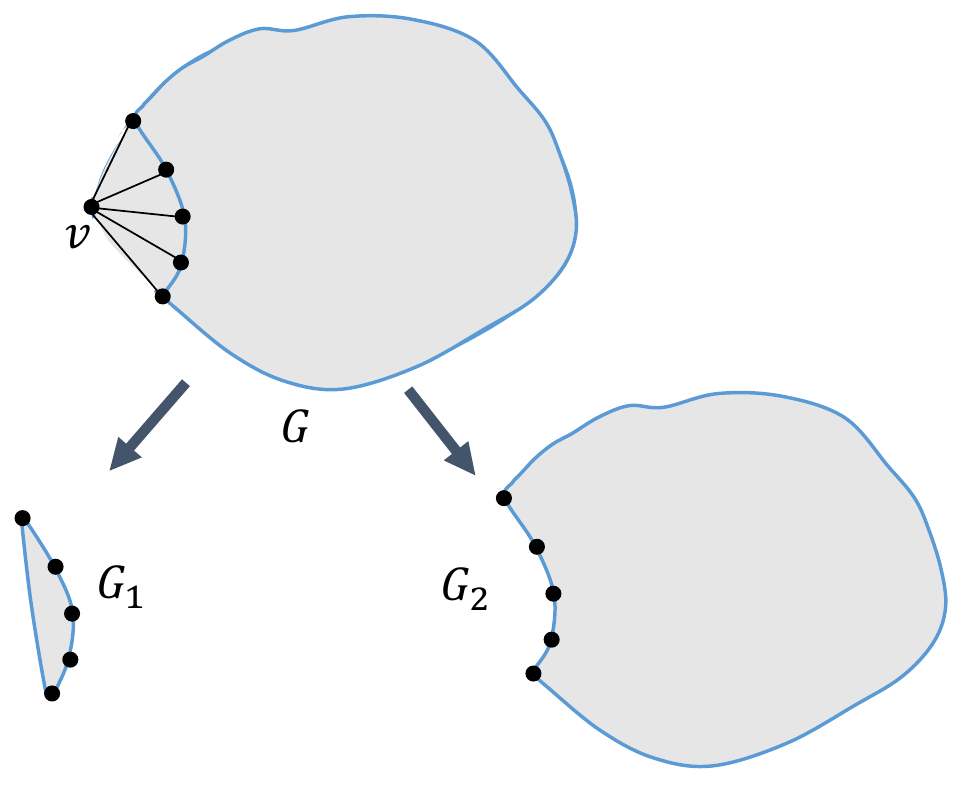}
  \caption{Illustration of the vertex splitting algorithm. $G_1$ contains all neighbors of $v$ (without $v$ itself) and the edges between them; $G_2$ contains all vertices and edges of $G$ except $v$ and the edges incident to $v$.}
  \label{fig:vertex_splitting}
\end{figure}

In \cite{Djidjev2017} we developed several algorithms for decomposing an input graph too large for DW into subgraphs small enough that the (Maximum Clique) QUBO corresponding to each generated subproblem fits DW. After solving each subproblem, the solutions are combined again into a solution to Maximum Clique for the original graph.

We describe here an improved version of the most universal of those algorithms, in the sense that it can be applied to all input graphs regardless of their structure. It works as follows. Given a graph $G=(V,E)$, pick any vertex $v$, e.g, with the smallest number of neighbors as in \cite{Djidjev2017}, and define two new graphs: a graph $G_1$ induced by all neighbors of $v$ (but without $v$ itself) and the graph $G_2$ induced by $V\setminus\{v\}$, that is $G$ without $v$ and its incident edges (Figure~\ref{fig:vertex_splitting}). Then find recursively the maximum cliques $C_1$  of $G_1$ and $C_2$ of $G_2$. If $|C_1|>|C_2|$, return as a maximum clique for $G$ the subgraph induced by $C_1\cup \{v\}$, otherwise return $C_2$. The recursion stops when $G$ is small enough to fit DW.

It is easy to see the recursion is finite, as both $G_1$ and $G_2$ contain no more than $|V|-1$ vertices each, since neither of them contains $v$, and that it correctly finds a clique of maximum size for $G$. Unfortunately, in the worst case it may generate an exponential number of subgraphs (exponential in the size of $G$). We will study in the next subsection how persistency analysis can help to reduce the number of subgraphs generated.

\subsection{Results with Qpbo}
In our experiments, we used the \textit{C++} package Qpbo of \cite{Rother2007,Kolmogorov2014} for finding persistencies, developed for computer vision applications. We access Qpbo through a Python interface that we adapted by  modifying the PyQpbo package \citep{JMLR:v15:mueller14a}. This section presents results on the effectiveness of the method for Maximum Clique. We investigate the application of Qpbo to QUBOs \eqref{eq:maxclique_acm} and \eqref{eq:maxclique_lucas} on random as well as special graphs (c-fat and Hamming graphs), both without and in connection with the graph splitting algorithm of Section~\ref{sec:decomp}. Moreover, we compare the two QUBO formulations \eqref{eq:maxclique_acm} and \eqref{eq:maxclique_lucas} presented in Section~\ref{sec:maxclique_qubo}.

\subsubsection{Qpbo on c-fat and Hamming graphs}
\label{sec:qpbo_hamming}

\begin{table}[t]
\centering\small{
  \begin{tabular}{|l|rr|rrr|}
  \hline
graph & n & q & strong & weak & probe \\ 
  \hline
c-fat & 200 &   1 & 0.00 & 0.00 & 100.00 \\ 
   & 200 &   5 & 0.00 & 0.00 & 100.00 \\ 
   & 500 &   1 & 0.00 & 0.00 & 100.00 \\ 
   & 500 &   5 & 0.00 & 0.00 & 100.00 \\ 
   \hline
   Hamming &   6 &   2 & 0.00 & 100.00 & 100.00 \\ 
   &   6 &   4 & 0.00 & 0.00 & 0.00 \\ 
   &   8 &   2 & 0.00 & 100.00 & 100.00 \\ 
   &   8 &   4 & 0.00 & 0.00 & 0.00 \\ 
   \hline
\end{tabular}}
\caption{Reduction in \% from strong and weak persistencies as well as probing on selected c-fat and Hamming graphs. Qubo for maximum clique.} 
\label{tab:maxclique}
\end{table}

We apply Qpbo to the QUBO formulation \eqref{eq:maxclique_acm}. The test graphs we use are from the \textit{1993 DIMACS Challenge on maximum cliques, coloring and satisfiabilty} \citep{johnson-trick-96} and have also been used in \cite{Boros2006}. Both graph families depend on two parameters: the number of vertices $n$ and an additional internal parameter, precisely the partition parameter $c$ for c-fat graphs and the Hamming distance $d$ for Hamming graphs. We use the generation algorithms of \cite{Hasselberg1993} for both graph families.

Table~\ref{tab:maxclique} shows the results (where both parameters $c$ and $d$ are summarized as a generic parameter $q$). We see that, in six out of eight cases, Qpbo can find persistencies for 100\% of the variables, while in the remaining two cases it cannot find any persistency. Also we observe that probing is the most effective algorithm, while finding strong persistencies is the least effective.

\newcommand{\newwidth}{0.86\columnwidth}
\begin{figure*}
  \centering
  \includegraphics[width=0.45\textwidth]{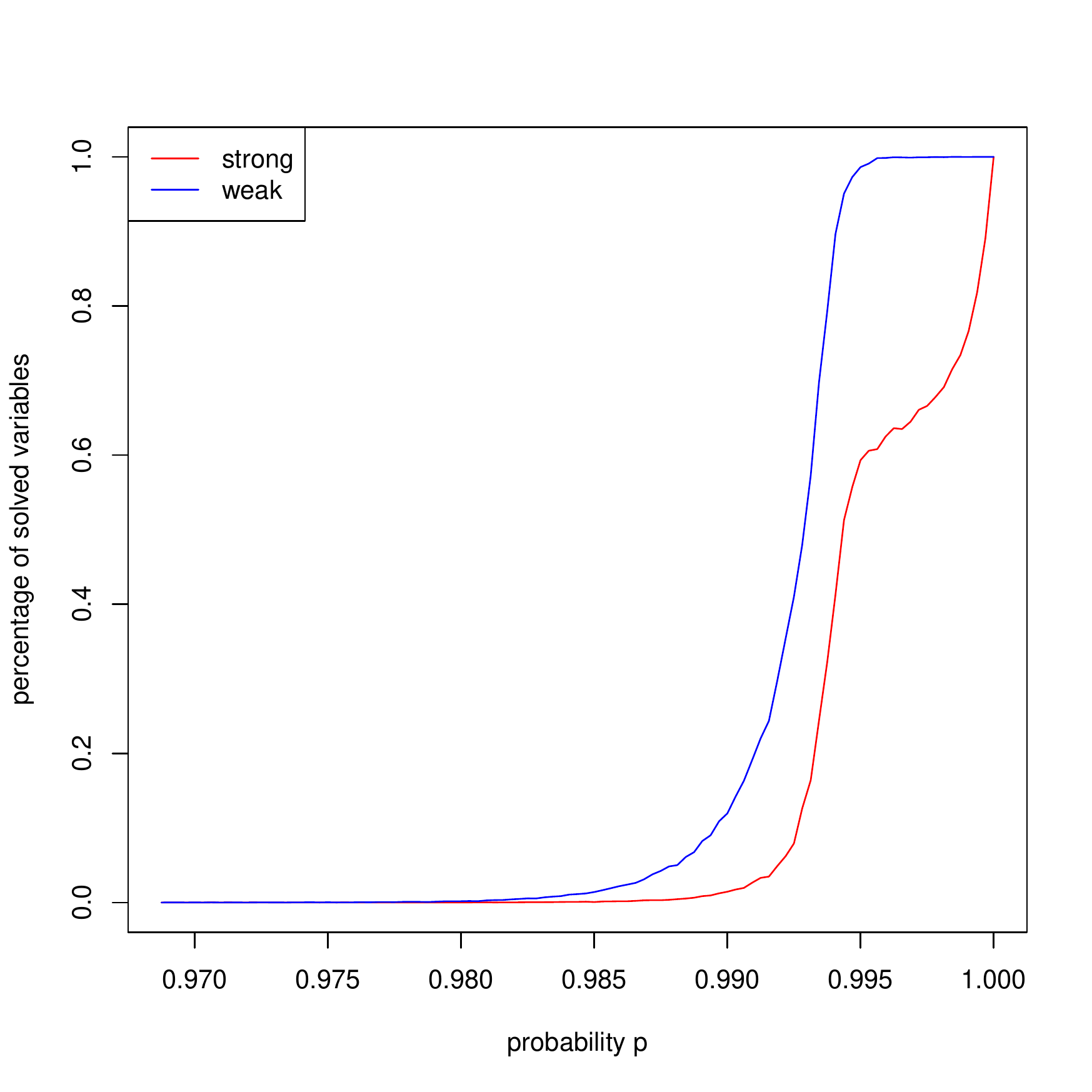}~
  \includegraphics[width=0.45\textwidth]{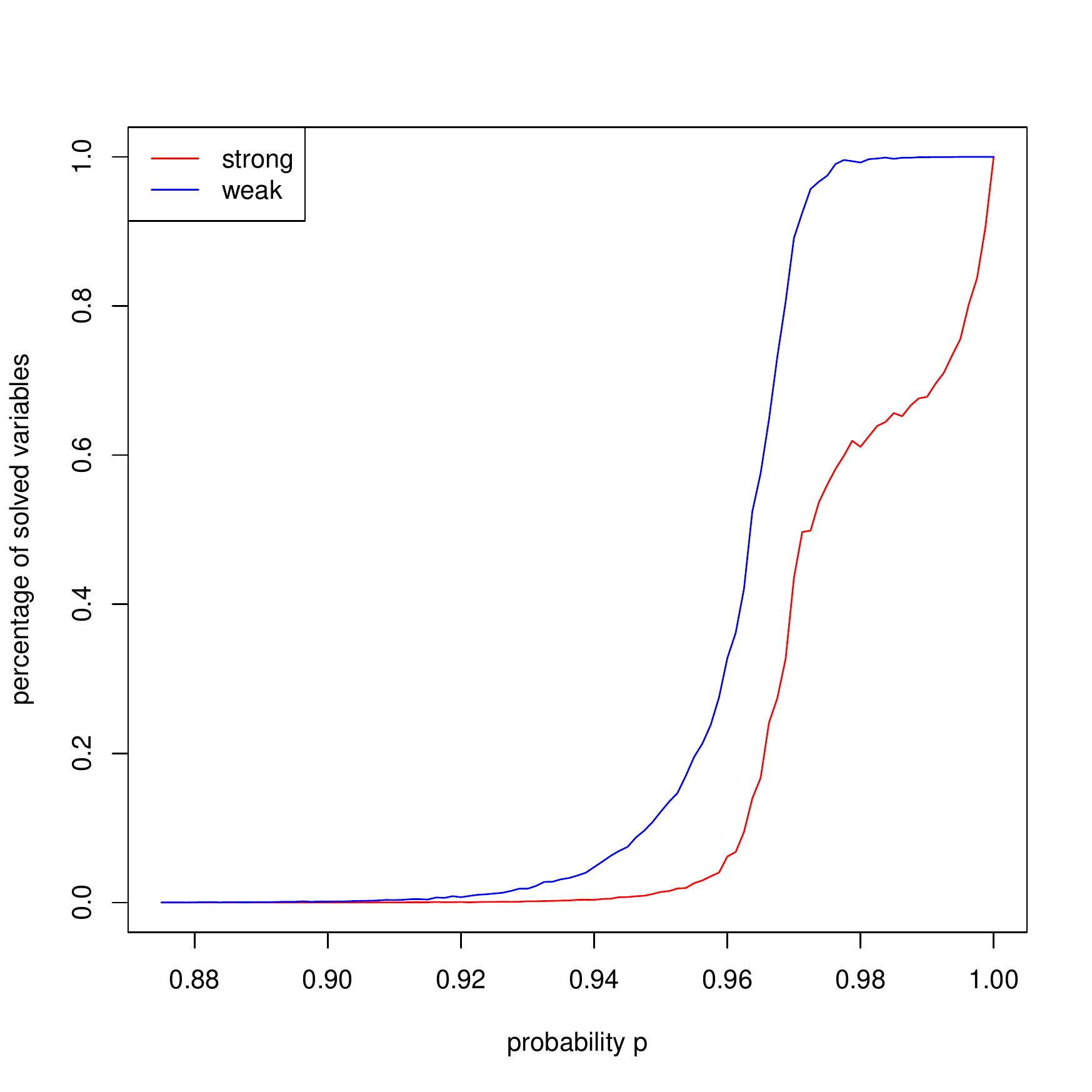}
  \caption{Percentage of found (strong and weak) persistencies with Qpbo as a function of the edge probability $p$
  with which random edges are inserted.
  C-fat graph with $n=500$ and $c=2$ (left) and Hamming graph with $n=8$ and $d=4$ (right).
  \label{fig:plot_fat500-2}}
\end{figure*}

In order to observe and compare a wider range of percentages of solved variables with Qpbo, we insert (or delete) edges from c-fat or Hamming graphs with a certain probability $p$ which we will vary.

Figure~\ref{fig:plot_fat500-2} (left) shows that when adding random edges (with edge probability $p$) to the $(500,2)$ c-fat graph, no strong and weak persistencies are found for both the original graph as well as for almost all dense modified graphs (since the x-axis starts at $p=0.97$). Persistencies (as a percentage of solved variables out of the $n=500$ variables in the QUBO) are only observed for $p \geq 0.97$ and increase to $100\%$ over a very short range. As expected, weak persistencies are found (slightly) earlier than strong ones. The behavior for the Hamming graph with $n=8$ and Hamming distance $d=4$ is similar.

\begin{figure*}
  \centering
  \includegraphics[width=0.45\textwidth]{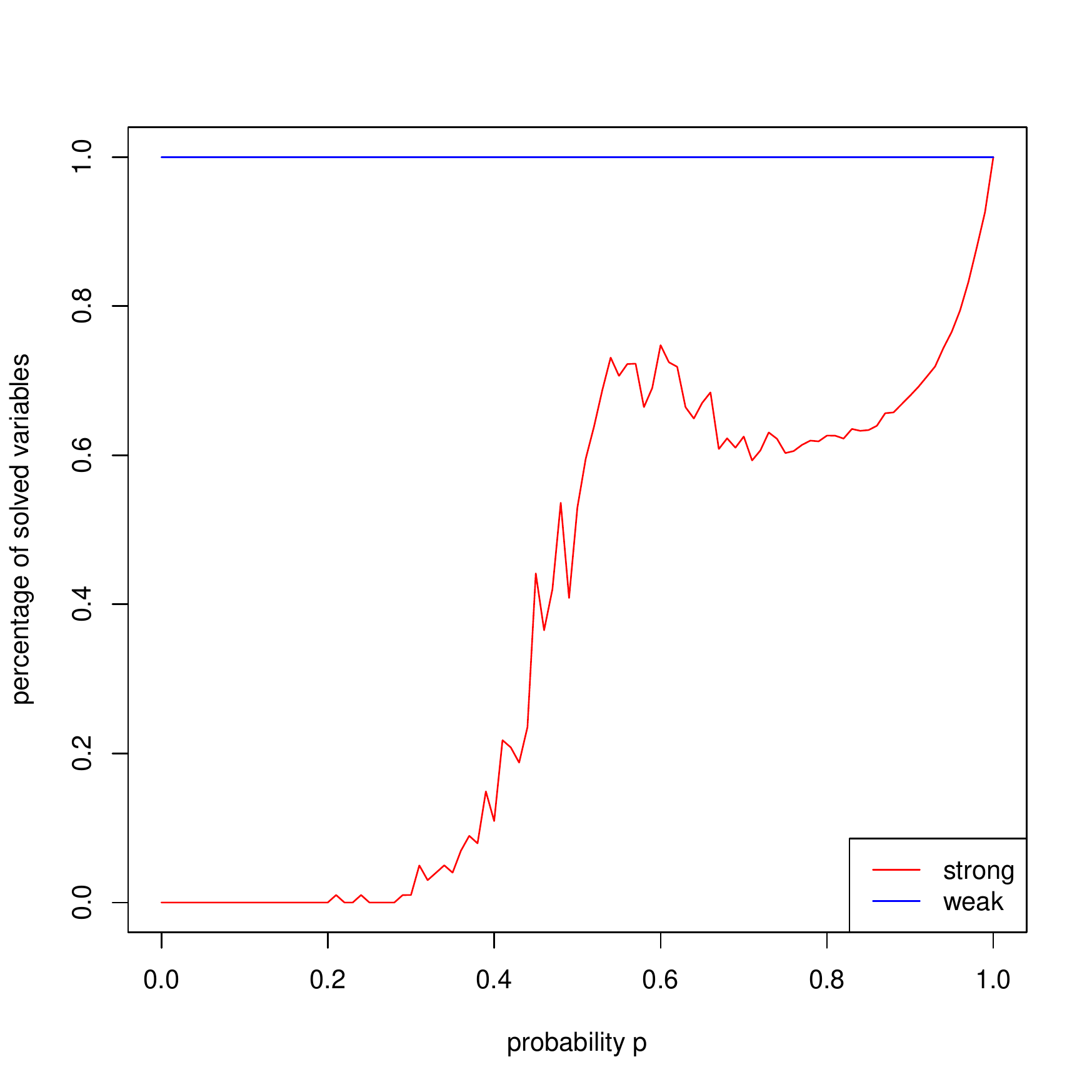}~
  \includegraphics[width=0.45\textwidth]{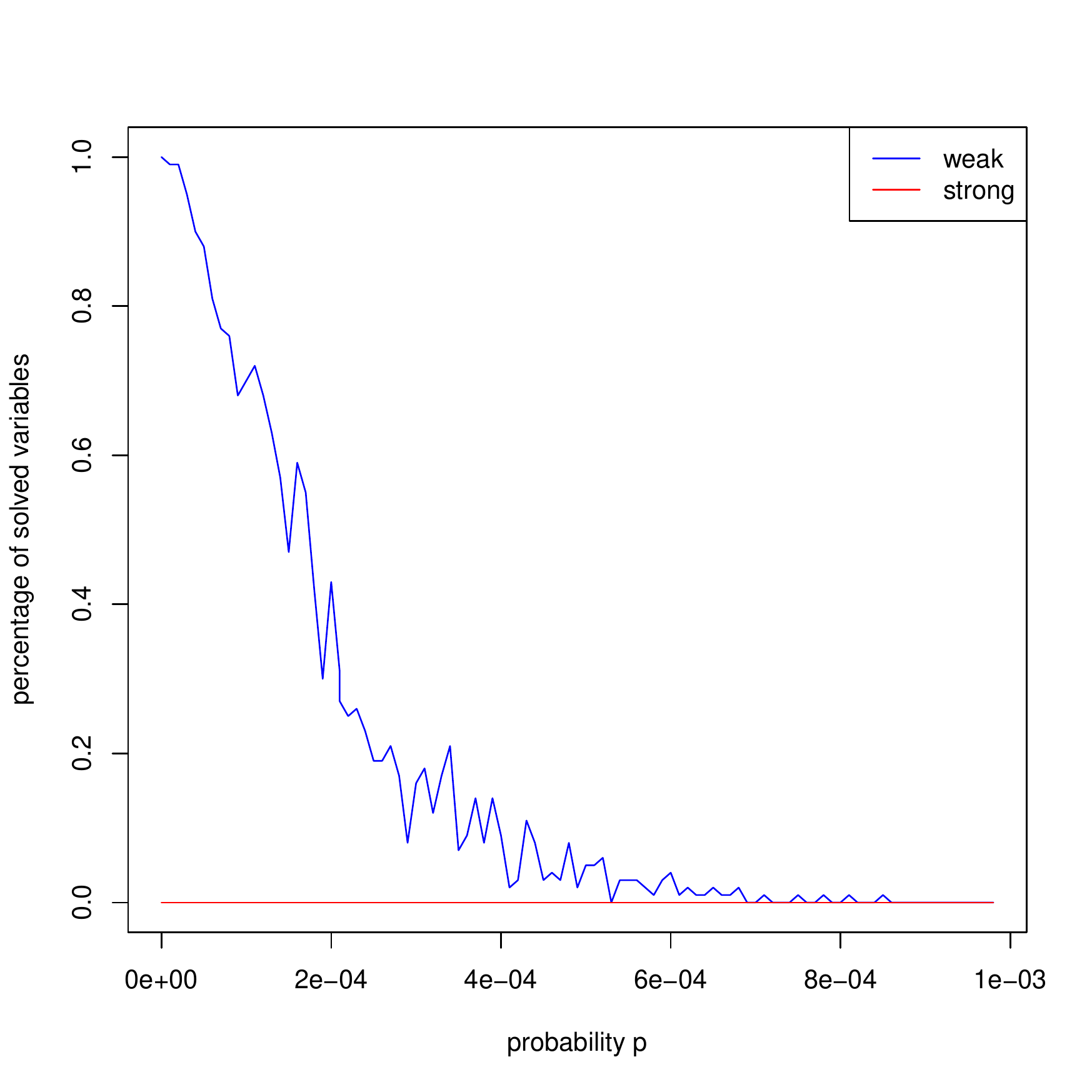}
  \caption{Percentage of found (strong and weak) persistencies with Qpbo as a function of the edge probability $p$
  with which random edges are inserted (in case of strong persistencies, left plot) or deleted (in case of weak persistencies, right plot).
  Hamming graph with $n=8$ and $d=2$.
  \label{fig:plot_ham8-2}}
\end{figure*}

Similarly to Figure~\ref{fig:plot_fat500-2}, Figure~\ref{fig:plot_ham8-2} shows strong and weak persistencies as a function of $p$ for the Hamming graph with $n=8$ and $d=4$. However, the behavior of strong and weak persistencies is different in this case. Figure~\ref{fig:plot_ham8-2} shows that Qpbo is able to assign a value to $100\%$ of all variables in the QUBO of the original Hamming graph via weak persistencies (see Table~\ref{tab:maxclique}). Nevertheless, strong persistencies are not found. We therefore add edges with probability $p$ to observe a progression of strong persistencies (Fig.~\ref{fig:plot_ham8-2}, left) and remove edges with probability $p$ for weak persistencies (Fig.~\ref{fig:plot_ham8-2}, right).

Figure~\ref{fig:plot_ham8-2} shows that strong persistencies increase gradually  over the entire interval $p \in [0,1]$, whereas the found weak persistencies disappear almost instantly as soon as edges are removed from the graph.

\subsubsection{Comparison of alternative QUBO formulations for Maximum Clique}
The QUBO formulation \eqref{eq:maxclique_acm} was used for all previous investigations. This is due to the fact that, in contrast to \eqref{eq:maxclique_lucas}, the former does not need the (usually unknown) clique size $K$ as an input parameter. Nevertheless, this section investigates whether the choice of formulation has an effect on the persistencies found and whether changing the QUBO formulation can result in more or fewer persistencies. If this is the case, then it may be worth investing time to search for formulations that are more suitable for the persistency algorithm.

In order to find the clique size that is needed for \eqref{eq:maxclique_lucas}, we first run an alternative algorithm to find the maximum clique size $K$ for each tested graph. We then compute the QUBO coefficients for \eqref{eq:maxclique_lucas} with parameter $K$ and identify strong and weak persistencies using Qpbo.

Table~\ref{tab:qubo_comparison} shows results for a Hamming graph with parameters $n=8$ and $d=2$ and a c-fat graph with parameters $n=200$ and $c=1$. As can be seen from the table, formulation \eqref{eq:maxclique_lucas} results in much larger QUBO problems (which are likely to also have a more complex structure, thus being more difficult to solve for Qpbo) and thus in fewer persistencies compared to \eqref{eq:maxclique_acm}, so clearly the latter formulation is preferable, at least for the type of graphs tested.

\begin{table}[t]
\centering\small{
\begin{tabular}{|c|c|c|c|c|c|c|}
  \hline
  &  & $p$  & formu- & QUBO  &  &   \\ 
  graph &  $n$   & or $c$ & lation &  size & strong & weak  \\ 
  \hline
  Ham- & $8$ & $2$ & \eqref{eq:maxclique_acm} & 1280 & $0$ & $100$ \\ 
  
  ming & $8$ & $2$ & \eqref{eq:maxclique_lucas} & 65536 & $0$ & $0$ \\ 
  \hline
  c-fat & $200$ & $1$ & \eqref{eq:maxclique_acm} & 481 & $69$ & $100$ \\ 
  
      & $200$ & $1$ & \eqref{eq:maxclique_lucas} & 40000 & $0$ & $0$ \\ 
  \hline
\end{tabular}}
\caption{Reduction of the number of variables in percentage for the two QUBO formulations for maximal clique on Hamming and fat graphs. } 
\label{tab:qubo_comparison}
\end{table}

\subsubsection{Combining Qpbo with the decomposition algorithm}
\label{sec:splitting_random}
We apply the graph splitting algorithm described in Section~\ref{sec:decomp} to solve the Maximum Clique problem on graphs of various sizes. However, before splitting a graph, we look for persistencies (in particular, for strong and weak persistencies since applying probing in each time step took prohibitively long) and, if found, reduce the size of the graph by removing the corresponding variables. We stop splitting when all generated (sub-)graphs are of size small enough to fit DW, which in our case is $45$ vertices.

We test the resulting algorithm on random graphs with $n=500$ vertices and an expected number of edges in the interval $[10000,40000]$.

\begin{figure}
\centering
  \includegraphics[width=0.5\textwidth]{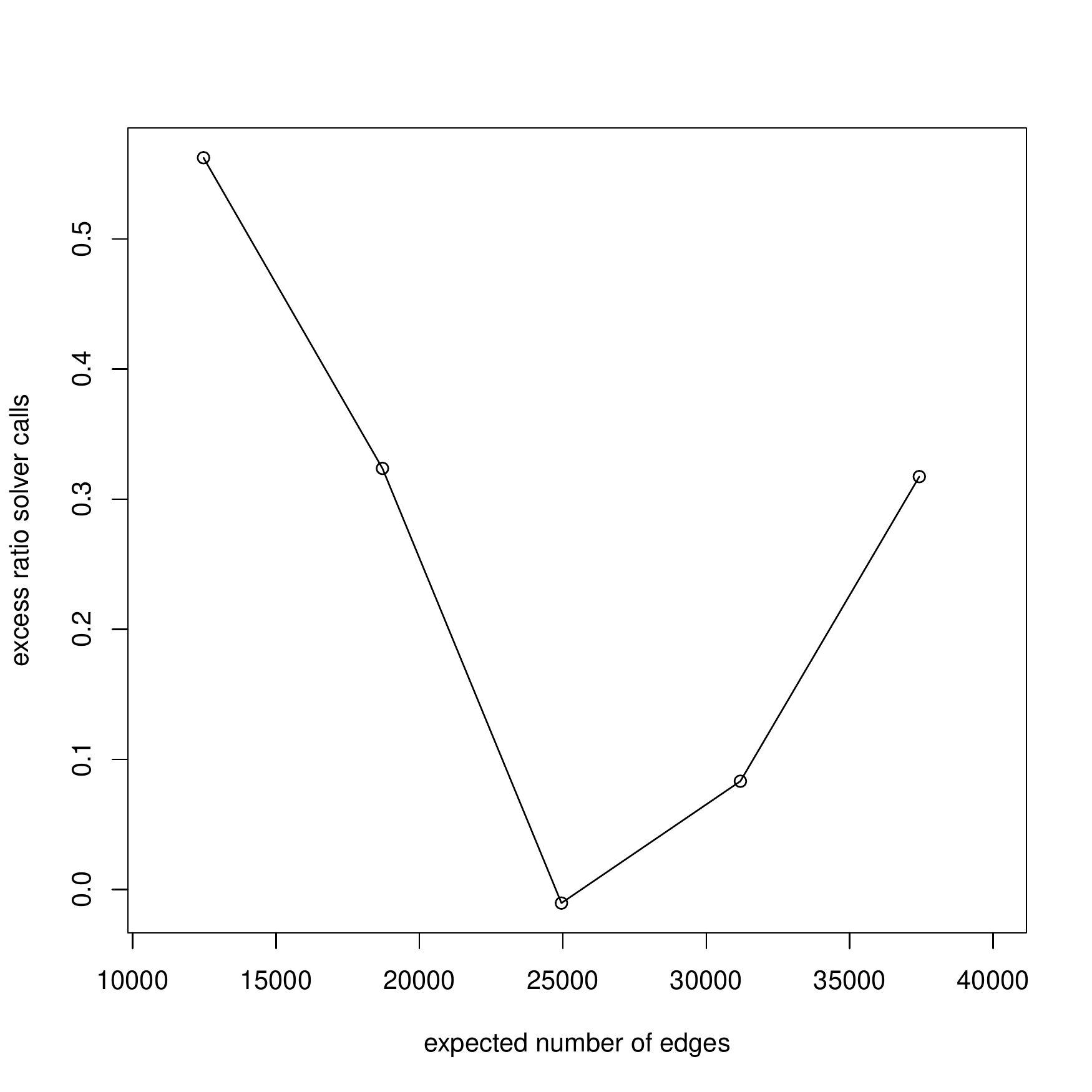}
  \caption{Proportional saving as a function of the expected number of edges of the random graph when using Qpbo in the graph splitting routine,
  computed as $(n_\text{qpbo}-n_\text{no-qpbo})/n_\text{no-qpbo}$,
  where $n_\text{qpbo}$ and $n_\text{no-qpbo}$ denote the number of solver calls when employing graph splitting with and without Qpbo.
  \label{fig:plot_splitting_ratio}}
\end{figure}

Figure~\ref{fig:plot_splitting_ratio} visualizes the proportional reduction in the number of generated subproblems. For this, we plot the ratio $(n_\text{qpbo}-n_\text{no-qpbo})/n_\text{no-qpbo}$ as a function of $p$, where $n_\text{qpbo}$ and $n_\text{no-qpbo}$ denote the number of solver calls when employing graph splitting with and without Qpbo, respectively. The figure shows that, especially for sparse graphs, substantial reductions with respect to the number of generated subproblems can be achieved.

\section{The maximum cut problem}
\label{sec:maxcut}
This section takes a closer look at the Maximum Cut problem introduced in Section~\ref{sec:intro}. We again state a QUBO formulation for Maximum Cut and present simulation results on $g$ and $U$ graphs, which have already been defined and previously been used for benchmarking Maximum Cut problems in \cite{Kim2001} and \cite{Boros2006}.

\subsection{QUBO and Ising formulation}
We use the QUBO formulation for Maximum Cut of \cite{BorosHammer1991} who show that the value $f$ of any cut in a graph $G=(V,E)$ is given as
\begin{align}
\label{eq:maxcut_QUBO}
f(x) = \sum_{(u,v)\in E} \left( x_u (1-x_v) + (1-x_u) x_v \right),
\end{align}
where $x_u \in \{0,1\}$ for all $u \in V$ is the indicator signalling which side of the cut vertex $u$ belongs to. Maximizing \eqref{eq:maxcut_QUBO} over $x \in \{0,1\}^{|V|}$ leads to a maximum cut.

The formulation \eqref{eq:maxcut_QUBO} can be simplified when switching the input range of $x_u$ to an Ising model. In \eqref{eq:maxcut_QUBO}, each term $x_u (1-x_v) + (1-x_u) x_v$ is zero if and only if $u$ and $v$ belong to the same side of the cut, and one otherwise. In an Ising formulation, that is when $x_u \in \{-1,+1\}$ for all $u \in V$, the same term can be expressed as $(1-x_u x_v)/2$, leading to the Hamiltonian
$$H_0(x) = \sum_{(u,v)\in E} \frac{1}{2} (1-x_u x_v),$$
which has to be maximized (just like \eqref{eq:maxcut_QUBO}). Equivalently, we can minimize
\begin{align}
H(x) &= - \sum_{(u,v)\in E} (1-x_u x_v)= -|E| + \sum_{(u,v)\in E} x_u x_v \nonumber  \\
&\sim \sum_{(u,v)\in E} x_u x_v,\label{eq:maxcut_ising}
\end{align}
since $|E|$ is constant. In the following we will solely use the simpler Ising formulation \eqref{eq:maxcut_ising}.

\subsection{Ising on \texorpdfstring{$g$}{Lg} and \texorpdfstring{$U$}{Lg} graphs}

\begin{table}[t]
\centering\small{
\begin{tabular}{|c|rr|rrr|}
  \hline
graph & n & p & strong & weak & probe \\ 
  \hline
g  & 500 & 2.50 & 0.00 & 13.40 & 25.70 \\ 
graphs   & 500 & 5.00 & 0.00 & 100.00 & 100.00 \\ 
   & 500 & 10.00 & 0.00 & 100.00 & 100.00 \\ 
   & 1000 & 2.50 & 0.00 & 11.40 & 25.60 \\ 
   & 1000 & 5.00 & 0.00 & 100.00 & 100.00 \\ 
   \hline
   U  & 500 & 5.00 & 0.00 & 1.60 & 6.20 \\ 
  graphs & 500 & 10.00 & 0.00 & 0.40 & 0.50 \\ 
   & 1000 & 5.00 & 0.00 & 2.70 & 5.60 \\ 
   & 1000 & 10.00 & 0.00 & 0.00 & 0.30 \\ 
   \hline
\end{tabular}}
\caption{Reduction in \% from strong and weak persistencies as well as probing on selected \textit{g} and \textit{U} graphs. Qubo for maximum cut.} 
\label{tab:maxcut}
\end{table}

We test Qpbo on the Ising formulation \eqref{eq:maxcut_ising} for the Maximum Cut problem applied to $g$ and $U$ graphs. Table~\ref{tab:maxcut} shows the results. As seen in the table, some  weak persistencies are found, but no strong ones. Unlike the results for Maximum Clique, where the found weak persistencies are either $0\%$ or $100\%$, here we observe more varied results. Again, probing is the most effective algorithm. Our results for Maximum Clique are roughly similar to the ones presented in \cite{Boros2006} (who, however, use a different and possibly more elaborate algorithm than Qpbo, which they did not make available).

Additionally, we again studied how the performance of Qpbo varies as the graph becomes denser by adding or removing random edges. These results are similar to the ones already described for Maximum Clique (Figures~\ref{fig:plot_fat500-2} and \ref{fig:plot_ham8-2}) and are thus not presented here.

\section{Conclusions}
\label{sec:conclusions}
This paper studies the use of preprocessing methods to reduce the size of QUBO and Ising integer programs. Such models have recently attracted increased attention since they allow constant time solutions on newly available quantum annealers. However, the small number of available qubits on such devices severely limits the range of solvable problems. This motivates the use of preprocessing methods to reduce the number of required qubits, thus extending the application range of such devices.

We use the package Qpbo \citep{Kolmogorov2014,JMLR:v15:mueller14a} to compute strong and weak persistencies in integer programs and test the suitability of preprocessing methods on two important NP-hard graph problems, the Maximum Clique and the Maximum Cut problems.

Our main findings can be summarized as follows:
\begin{enumerate}
  \item The observed reduction in variables seems to be very problem-specific, both in terms of the QUBO/ Ising formulation considered as well as the employed test graphs. Test graphs on which Qpbo achieved $100\%$ reductions for Maximum Clique did not lead to any reduction for Maximum Cut or vice versa. We observe $0\%$ and $100\%$ reductions considerably more often than reductions between those extremes.
  \item When applied to a decomposition algorithm which divides up a maximum clique computation on graphs of arbitrary size into many smaller subproblems, Qpbo can be used to reduce the number of generated subproblems. This reduction seems especially significant for sparser graphs.
  \item Using two different QUBO formulations for Maximum Clique, we show that the performance of Qpbo is not only dependent on the type of problem solved and the input graph, but also on the specific QUBO formulation used. This, therefore, justifies research into finding formulations for different optimization problems that are more suitable for identifying persistencies. This is consistent with the much more general observation in combinatorial optimization practice that the choice of formulation for a given optimization problem can have a huge effect on the time it will take to solve the resulting problem using popular solvers.
\end{enumerate}

The effectiveness of Qpbo varied widely between seemingly similar problem instances. An especially interesting and practically important open problem is to be able to characterize input graphs and problem formulations that are more suitable for identification of persistencies.

Another interesting open problem is how to best speed up the persistency finding algorithms. For instance, in the case where persistency finding (via Qpbo) is combined with the decomposition algorithm (for Maximum Clique), we currently run Qpbo on every subgraph generated during the graph splitting process. A better way is to try to use information collected during the persistency analysis for the larger graph in order to infer persistencies for the smaller ones, rather than computing them from scratch, thereby saving computation time.


\begin{thebibliography}{}
\bibitem[Balas and Yu, 1986]{Balas1986}
Balas, E. and Yu, C. (1986).
\newblock Finding a maximum clique in an arbitrary graph.
\newblock {\em SIAM J Comp}, 15:1054--1068.

\bibitem[Boros and Hammer, 1991]{BorosHammer1991}
Boros, E. and Hammer, P. (1991).
\newblock {The max-cut problem and quadratic 0-1 optimization; Polyhedral
  aspects, relaxations and bounds}.
\newblock {\em Ann Oper Res}, 33:151--180.

\bibitem[Boros and Hammer, 2001]{BorosHammer2001}
Boros, E. and Hammer, P. (2001).
\newblock {Pseudo-Boolean Optimization}.
\newblock {\em Rutcor Report}, pages 1--83.

\bibitem[Boros et~al., 2006]{Boros2006}
Boros, E., Hammer, P., and Tavares, G. (2006).
\newblock {Preprocessing of Unconstrained Quadratic Binary Optimization}.
\newblock {\em Rutcor Research Report}, RRR 10-2006:1--58.

\bibitem[Bunyk et~al., 2014]{Bunyk2014}
Bunyk, P., Hoskinson, E., Johnson, M., Tolkacheva, E., Altomare, F., Berkley,
  A., Harris, R., Hilton, J., Lanting, T., Przybysz, A., and Whittaker, J.
  (2014).
\newblock Architectural considerations in the design of a superconducting
  quantum annealing processor.
\newblock {\em IEEE Trans on Appl Superconductivity}, 24(4):1--10.

\bibitem[{D-Wave}, 2016]{dwave2016}
{D-Wave} (2016).
\newblock Introduction to the {D-Wave} quantum hardware.

\bibitem[Djidjev et~al., 2017]{Djidjev2017}
Djidjev, H., Chapuis, G., Hahn, G., and Rizk, G. (2017).
\newblock Finding maximum cliques on a quantum annealer.
\newblock {\em ACM Computing Frontiers}, 1(1):1--8.

\bibitem[Hammer et~al., 1984]{Hammer1984}
Hammer, P., Hansen, P., and Simeone, B. (1984).
\newblock {Roof duality, complementation and persistency in quadratic 0-1
  optimization}.
\newblock {\em Mathematical Programming}, 28(2):121--155.

\bibitem[Hasselberg et~al., 1993]{Hasselberg1993}
Hasselberg, J., Pardalos, P., and Vairaktarakis, G. (1993).
\newblock {Test Case Generators and Computational Results for the Maximum
  Clique Problem}.
\newblock {\em Journal of Global Optimization}, 3:463--482.

\bibitem[Johnson and Trick, 1996]{johnson-trick-96}
Johnson, D. and Trick, M., editors (1996).
\newblock {\em Clique, Coloring, and Satisfiability: Second {DIMACS}
  Implementation Challenge, {DIMACS}}, volume~26.
\newblock American Mathematical Society.

\bibitem[Johnson et~al., 2011]{Johnson2011}
Johnson, M., Amin, M., Gildert, S., Lanting, T.~Hamze, F., Dickson, N., Harris,
  R., Berkley, A., Johansson, J., Bunyk, P., Chapple, E., Enderud, C., Hilton,
  J., Karimi, K., Ladizinsky, E., Ladizinsky, N., Oh, T., Perminov, I., Rich,
  C., Thom, M., Tolkacheva, E., Truncik, C., Uchaikin, S., Wang, J., B., W.,
  and Rose, G. (2011).
\newblock Quantum annealing with manufactured spins.
\newblock {\em Nature}, 473:194--198.

\bibitem[Kim et~al., 2001]{Kim2001}
Kim, S.-H., Kim, Y.-H., and Moon, B.-R. (2001).
\newblock {A Hybrid Genetic Algorithm for the MAX CUT Problem}.
\newblock {\em Proceeding GECCO'01 Proceedings of the 3rd Annual Conference on
  Genetic and Evolutionary Computation}, pages 416--423.

\bibitem[King et~al., 2015]{King2015}
King, J., Yarkoni, S., Nevisi, M.~M., Hilton, J.~P., and McGeoch, C.~C. (2015).
\newblock Benchmarking a quantum annealing processor with the time-to-target
  metric.
\newblock {\em arXiv:1508.05087}, pages 1--29.

\bibitem[Kolmogorov, 2014]{Kolmogorov2014}
Kolmogorov, V. (2014).
\newblock {QPBO v1.4 software package}.
\newblock {\em C++ package manual}.

\bibitem[Lucas, 2014]{Lucas2014}
Lucas, A. (2014).
\newblock Ising formulations of many np problems.
\newblock {\em Frontiers in Physics}, 2(5):1--27.

\bibitem[M{\"u}ller and Behnke, 2014]{JMLR:v15:mueller14a}
M{\"u}ller, A.~C. and Behnke, S. (2014).
\newblock {Pystruct - Learning Structured Prediction in Python}.
\newblock {\em Journal of Machine Learning Research}, 15:2055--2060.

\bibitem[Rother et~al., 2007]{Rother2007}
Rother, C., Kolmogorov, V., Lempitsky, V., and Szummer, M. (2007).
\newblock {Optimizing Binary MRFs via Extended Roof Duality}.
\newblock {\em Proceedings CVPR, IEEE Computer Society Conference on Computer
  Vision and Pattern Recognition}, pages 1--15.
\end{thebibliography}

\end{document}